\documentclass[journal]{IEEEtran} 

\usepackage{amsmath, amssymb, amsfonts, amsbsy, mathrsfs, bm}
\usepackage{cite, enumerate}
\usepackage[table]{xcolor}
\usepackage[normalem]{ulem}
\usepackage{dsfont} 

\newtheorem{remark}{Remark}
\newtheorem{assumption}{Assumption}
\newtheorem{defn}{Definition} 
\newtheorem{prop}[defn]{Proposition}

\def\diag{{\mathrm{diag}}}

\newcommand{\E}[1]{\mathbb{E}\left[ #1 \right]}

\def\bw{{\pmb{w}}}
\def\bq{{\pmb{q}}}
\def\bu{{\pmb{u}}}
\def\bd{{\pmb{d}}}
\def\bv{{\pmb{v}}}
\def\bh{{\pmb{h}}}
\def\bp{{\pmb{p}}}
\def\br{{\pmb{r}}}
\def\bh{{\pmb{h}}}
\def\bz{{\pmb{z}}}
\ifCLASSINFOpdf
   \usepackage[pdftex]{graphicx}
 \else
 
\fi

\begin{document}
%
\title{Performance Analysis of Incremental LMS over Flat Fading Channels}

\author{Azam~Khalili,
        Amir~Rastegarnia
        
}



\maketitle

\begin{abstract}
We study the effect of fading in the communication channels between sensor nodes on the performance of the incremental least mean square (ILMS) algorithm, and derive steady state performance metrics, including the mean-square deviation (MSD), excess mean-square error (EMSE) and mean-square error (MSE). We obtain conditions for mean convergence of the ILMS algorithm, and show that in the presence of fading channels, the ILMS algorithm is asymptotically biased. Furthermore, the dynamic range for mean stability depends only on the mean channel gain, and under simplifying technical assumptions, we show that the MSD, EMSE and MSE are non-decreasing functions of the channel gain variances, with mean-square convergence to the steady states possible only if the channel gain variances are limited. We derive sufficient conditions to ensure mean-square convergence, and verify our results through simulations.

\end{abstract}


%


\section{Introduction}

Many applications require us to learn or estimate some parameters related to a phenomenon of interest, based on all the available observations in a network \cite{Akyildiz02,Akyildiz2005,Bulusu2005,Chong03,tu12}. Various schemes have been investigated in the literature to facilitate this, including the use of a fusion center to perform decentralized detection and estimation \cite{Tay2008,Xiao05,Aysal08}, and distributed methods that do not rely on a fusion center. An example of a distributed method is the consensus strategy \cite{Boyd06,Schizas09,Kar11} in which each node performs a local estimation and fuses its estimate with those of its neighbors so that all nodes converge to the same estimate as the number of iterations increases. An alternative distributed approach is the incremental update method \cite{lopes07, sayed06, li10, tak08, lopes10, catt11b}, which relies on nodes passing updates to each other in a Hamiltonian cycle in the network. Yet another method is the diffusion strategy \cite{lopes08,catt08,catt09,tak09,tu11,chen12,sayed13a,sayed13b} that performs online estimation in a distributed manner by letting each node update its local estimate based on information from all its neighbors and over multiple observation epochs. In both the incremental update and diffusion approaches, techniques from distributed optimization \cite{Tsitsiklis86,Nedic2009} are incorporated into the updates and local estimates need not converge to the same value, which leads to better performance over consensus strategies \cite{Tu12a}. Networks that rely on in-network processing at each node while allowing the node estimates to update and adapt to new sensor observations have come to be known as adaptive networks \cite{lopes07,tu11,chen12,sayed13a}.

One of the most popular approaches to modeling the underlying process observed by the nodes in an adaptive network is to adopt a linear model and use a least mean squares (LMS) criterion in the estimation procedure. This is because of its simplicity and wide applicability, for example in localization of targets, collaborative spectrum sensing, and in modeling group behaviors in biological systems \cite{sayed13a,tu12}. In this paper, we investigate the performance of the incremental LMS (ILMS) algorithm in a wireless adaptive network with communication links between neighboring nodes modeled as fading channels. In the original ILMS strategy proposed by \cite{lopes06}, it is assumed that nodes communicate with each other via ideal links. However, this is typically not true in practice. In \cite{azam11a,azam11b,azam12a,azam12b,zhao12a}, the effect of additive link noise in the communication channels between nodes have been investigated. In \cite{zhao12}, the performance of general adaptive diffusion algorithms in the presence of imperfect information exchanges, including quantization errors, and model non-stationarities has been considered. All these works however do not apply directly to a wireless sensor network, whose communication links are usually modeled by fading channels \cite{Tse05}. The references \cite{Abdolee11,Abdolee13} propose diffusion LMS algorithms for wireless sensor networks with fading channels but under the assumption that channel state information is known so that channel equalization can be performed. However, in practice the fading coefficients can only be estimated up to an uncertainty. Moreover, such channel equalization may be impractical for channels with time varying channel information. As wireless sensor networks have many important practical applications like building structure monitoring, it is therefore important to investigate the performance of adaptive networks in the presence of fading channels, without full knowledge of channel state information. 

In this paper, our objective is to investigate how the performance of the ILMS algorithm is affected by the channel fading statistics of the communication channels between nodes. We restrict our analysis to the ILMS algorithm instead of the more general class of diffusion algorithms because of the significantly more complex analysis involved for diffusion algorithms, which is out of the scope of this paper. Nevertheless, to the best of our knowledge, this work is the first to analytically evaluate the performance of the ILMS algorithm in a practical wireless sensor network with fading channels, without exact knowledge of channel state information. We believe that many of the insights obtained in this paper are applicable to more general diffusion algorithms, which will be addressed as part of future work. Our main contributions in this paper are the following:
\begin{enumerate}[(i)]
	\item We show that the ILMS algorithm over fading channels, like the traditional ILMS algorithm, is stable in mean if the step size is chosen in an appropriate range. We show that this range now not only depends on the regression correlation, but also on the mean of the channel fading gains. Moreover, our analysis reveals that in general, fading communication channels lead to biased estimates in steady state.
	\item We derive closed-form expressions for the steady state performance metrics, including the mean-square deviation (MSD), excess mean-square error (EMSE) and mean-square error (MSE), of the ILMS algorithm in the presence of fading channels, and under a Gaussian model. We show explicitly how these metrics are affected by the fading channel statistics.
	\item We derive sufficient conditions for the convergence of the MSD, EMSE and MSE, and show that for a fixed step size, mean-square stability is lost if the channel gain variances become large. Under simplifying technical conditions, we show that under the decibel (dB) scale, the MSD, EMSE and MSE are approximately non-decreasing linear functions of the channel gain second order moments.
\end{enumerate}
We also perform extensive simulations to verify that our theoretical analysis closely matches the actual steady state performance observed in a network.

The remainder of this paper is organized as follows. In Section~\ref{sect:ILMS}, we describe our system model and assumptions. In Section~\ref{sect:Performance}, we present theoretical analysis of the steady state performance of the ILMS algorithm over fading channels, and discuss some of our results. In Section~\ref{sect:Simulation}, we present simulation results to verify our theoretical analysis, and we conclude in Section~\ref{sect:Conclusion}.

\textit{Notation}: We adopt boldface letters for random quantities. The symbol $^*$ denotes conjugation for scalars and Hermitian transpose for matrices. The notation $\rm{col}\{\cdot\}$ denotes a column vector (or matrix) with the specified entries stacked on top of each other. The notation $\rm{diag}\{\cdot\}$ will be used in two ways: $X=\mathrm{diag}\{x\}$ is a diagonal matrix whose entries are those of the vector $x$, and $x=\rm{diag}\{X\}$ is a vector containing the main diagonal of $X$. The exact meaning of this notation will be clear from the context. We let $\mathds{1} \triangleq {\rm{diag}}\{ I\}$ denote a vector consisting of all 1s. We use $\lambda_{\max}(A)$, $\lambda_{\min}(A)$ and $\rho(A)$ to denote the largest eigenvalue, smallest eigenvalue, and spectral radius of $A$, respectively. If $\Sigma$ is a matrix, we use the notation $\| x\|_{\Sigma}^2=x^* \Sigma x$ for the weighted square norm of $x$. If $\sigma$ is a vector, the notation $\| x\|_{\sigma}$ is used to represent $\| x \|_{\mathrm{diag}\{\sigma\}}$.

\section{Problem Formulation}\label{sect:ILMS}

In this section, we first describe the data model, and give a brief overview of the traditional ILMS algorithm in \cite{lopes06}. We then describe how the ILMS updates are changed when information is communicated over fading channels. Finally, we list the assumptions we are making throughout this paper.

Consider a network composed of $N$ nodes. At time $i$, node $k$ observes a scalar measurement $\bd_k(i)$ and a $1\times M$ regression vector $\bu_{k,i}$,  which are related via a linear regression model
\begin{equation} \label{dmodel}
\bd_k(i)=\bu_{k,i} w^o+\bv_k(i),
\end{equation}
where $\bv_k(i)$ is the observation (or measurement) noise, and the $M \times 1$ vector $w^o=[w^o(1), w^o(2), \cdots,w^o(M)]^T$ is a deterministic (but unknown) vector. Based on the intended application, $w^o$ may have different physical meanings. For example, $w^o$ may represent the location of a target in space, or the parameters of an AR model \cite{sayed13a, sayed13b}. We make the following assumptions regarding the data model in \eqref{dmodel}. These assumptions are commonly assumed in the literature \cite{lopes06,sayed13a, sayed13b}.
\begin{assumption}\label{assumpt:model}\
\begin{enumerate}[(i)]
	\item For $k=1,\ldots,N$ and $i \geq 1$, the regression vectors $\bu_{k,i}$ are independent over node indices $k$ and observation times $i$. 
	\item The measurement noises $\bv_k(i)$ for all nodes $k=1,\ldots,N$, and all observation times $i \geq 1$, are zero-mean and independent of each other and the regression vectors $\bu_{k,i}$.
\end{enumerate}
\end{assumption}

The goal of the network is to estimate $w^o$, at every node $k$, using all observed  data in the entire network. Mathematically, the network seeks to find $w^o$ that minimizes the following LMS objective function:
\begin{equation} \label{prob}
\sum_{k=1}^N \mathbb{E}\left[|\pmb{d}_k(i)-\pmb{u}_{k,i} w|^2\right].
\end{equation}
It can be shown that the minimizer $w^o$ of \eqref{prob} satisfies the normal equation \cite{sayed08}.
\begin{align*}
{w}^o &=\left(\sum_{k=1}^N R_{u,k} \right)^{-1} \left(\sum_{k=1}^N r_{du,k} \right), 
\end{align*} 
where 
\begin{align*} 
R_{u,k} &=\mathbb{E}[\bu_{k,i}^*\bu_{k,i}],\ \  \textrm{and}\ \ r_{du,k}=\mathbb{E}[\bd_k(i)\bu_{k,i}^*]. 
\end{align*} 

\subsection{The ILMS Algorithm over Ideal Links}
In \cite{lopes06, lopes07}, an adaptive network based on incremental cooperation amongst nodes at every observation time has been developed wherein nodes communicate through a pre-established Hamilton cycle. This is known as the ILMS algorithm. Each node $k$ receives a local estimate from the previous node $k-1$, updates it using its local data, and then sends it to the next node $k+1$, where node indices are modulo $N$. The update equations for the ILMS algorithm, at iteration $i$ is given by \cite{lopes07} 
\begin{equation} \label{ilms}
\left\{ \begin{array}{l}
 \bw _{0,i}  \leftarrow \pmb{w}_{N, i - 1}  \\ 
 \bw _{k,i}  = \mathop {\bw _{k - 1,i}  + \mu_k } \bu_{k,i}^* ( {\bd_k (i) - \bu_{k,i} \bw _{k - 1,i} })\, \\ 
 \end{array} \right.
\end{equation}
where $\bw_{k,i}$ is the local estimate of the node $k$ at time $i$. It is shown in \cite{lopes07, sayed13a} that as $i \rightarrow \infty$, we have ${\bw_{k,i}} \to {w^o}$ in the mean, for every node $k$, and for an appropriately chosen set of step sizes $\{\mu_k: k=1,\ldots,N\}$.


\subsection{The ILMS Algorithm over Fading Channels}
\begin{figure}[!t]	                                                             
\centering 
\includegraphics [width=8.5cm] {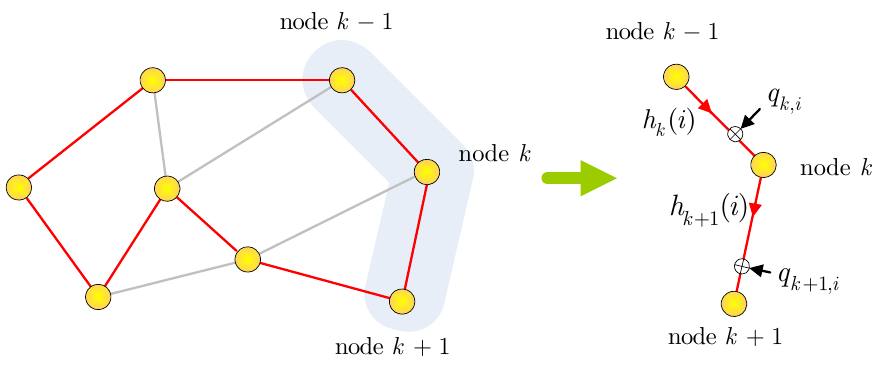} 
\centering \caption{The incremental LMS adaptive network with fading channels.}
 \label{fig:net}	
\end{figure}
In this paper, we consider the case where nodes communicate over fading channels (see Fig. \ref{fig:net}). To model model the Impact of channel, let denote by $\br_{k,i}$ the received signal at node $k$ and time $i$. By incorporating the impact of fading channels we have
\begin{align} \label{recsig}
\br_{k,i} &=\pmb{h}_k(i) \bw_{k - 1,i} + \bq_{k,i}
\end{align}
Using \eqref{recsig}, the update equation for node $k$ in \eqref{ilms} becomes
\begin{align} \label{ueinr}
\bw_{k,i} &=\br_{k,i}+ {\mu _k}{\pmb{u}}_{k,i}^ * \left({{\pmb{d}}_k}(i) - {{\pmb{u}}_{k,i}}\br_{k,i} \right)
\end{align}
where ${\pmb{h}_k}(i)$ is the channel gain at time $i$ for the communication channel between node $k-1$ and $k$, and $\bq_{k,i}$ is the additive channel noise. The update equation \eqref{ueinr} in terms of $\pmb{h}_k(i)$, $\bw_{k - 1,i}$ and $\bq_{k,i}$ as
\begin{align} 
\bw_{k,i} & = \pmb{h}_k(i) \bw_{k - 1,i} + {{\pmb{q}}_{k,i}} \nonumber \\ 
&\quad  + {\mu _k}{\pmb{u}}_{k,i}^ * \left({{\pmb{d}}_k}(i) - {{\pmb{u}}_{k,i}}({\pmb{h}_k}(i) \bw _{k - 1,i}+ {{\bq}_{k,i}}) \right), \label{films}
\end{align}
We make the following assumptions regarding the fading channel statistics.
\begin{assumption}\label{assumpt:fading}\
\begin{enumerate}
  \item We assume phase coherent reception at every node $k$ and model the channel gain as a non-negative random variable with $m_k = \E{\bh_k(i)}$ and $s_k = \E{\bh_k^2(i)}$ as the mean and second order moment of the channel gain for node $k$, respectively. 
	\item The channel gains $\pmb{h}_k(i)$ for all nodes $k=1,\ldots,N$, and all observation times $i \geq 1$, are independent of each other. For each node $k$, the channel gains $\{\bh_k(i) : i \geq 1\}$ are identically distributed. 
	\item The channel gains $\pmb{h}_k(i)$ for all nodes $k=1,\ldots,N$, and all observation times $i \geq 1$, are independent of $(\pmb{d}_l(j), \pmb{u}_{l,j})$ for all $l$ and $j$. 
	\item The additive channel noises $\bq_{k,i}$ for all nodes $k=1,\ldots,N$, and all observation times $i \geq 1$, are zero-mean and independent of each other, the observation noises $\bv_l(j)$, and the regression vectors $\bu_{l,i}$ for all $l$ and $j$. The channel noise $\bq_{k,i}$ has covariance matrix $Q_k=\mathbb{E}\left[\bq_{k,i}\bq_{k,i}^*\right]$.
\end{enumerate}
\end{assumption}

\begin{remark}
When the channel fading coefficients $\bh_{k}(i)$ are available, we can use zero-forcing (ZF) equalizer to mitigate the effects of fading channels. To show this, let denote by $\bz_{k,i}$ the ZF equalizer coefficient which is given by
\begin{align} \label{zfc}
\bz_{k}(i)&=\frac{\bh^*_{k}(i)}{\|\bh_{k}(i)\|^2}
\end{align}
Multiplying the received signal with the ZF equalizer coefficient gives
\begin{align} \label{impsig}
\br'_{k,i}&= \bz_{k}(i)\br_{k,i} =\bw_{k-1,i}+\bq'_{k,i}
\end{align}
where $\bq'_{k,i} \triangleq \frac{\bh^*_{k}(i)}{\|\bh_{k}(i)\|^2} \bq_{k,i}$ can be interpreted as the effect of fading channel in the special case when the fading coefficients $\bh_{k}(i)$ are available. We can see that in this case the ILMS algorithm with fading channels behaves like the ILMS network with noisy links. 
\end{remark}
\begin{remark}
In practice, the channel fading coefficients $\bh_{k}(i)$ are not available and only can be estimated by e.g. training data. In this case even if the nodes perform channel state estimation, it is not possible to measure the channel gains with absolute certainty, especially if the channels experience fast fading. To further the motivation for using \eqref{films}, let $\hat{\bh}_{k}(i)$ be the estimate of the fading coefficient $\bh_{k}(i)$. Then, the ZF equalizer coefficient becomes
\begin{align} \label{czfc}
\bz'_{k}(i)&=\frac{\hat{\bh}^*_{k}(i)}{\|\hat{\bh}_{k}(i)\|^2}
\end{align}
Using \eqref{czfc} $\br'_{k,i}$ in \eqref{impsig} changes to
\begin{align} \label{cimpsig}
\br'_{k,i}&= \bz'_{k}(i)\br_{k,i} =\bh'_k(i)\bw_{k-1,i}+\bq''_{k,i}
\end{align}
where
\begin{align} 
\bh'_k(i) \triangleq \frac{\bh_k(i) \hat{\bh}^*_k(i)}{\|\hat{\bh}_{k}(i)\|^2},\ \ \bq''_{k,i} \triangleq \frac{\hat{\bh}^*_{k}(i)}{\|\hat{\bh}_{k}(i)\|^2} \bq_{k,i}
\end{align}
We can see from \eqref{cimpsig} that when the nodes perform channel state estimation the channel gain for node $k$ can still be modeled as a non-negative random variable with a non-trivial variance. Of particular interest is the case where channel estimation is sufficiently accurate on average so that the channel gain can be partially accounted for at the receiver node. In this case, we can assume that $m_k=1$ and $s_k > 1$. In the next section, we show how the performance of the ILMS algorithm depends on both $m_k$ and $s_k$.
\end{remark}

\section{Performance Analysis}\label{sect:Performance}

In this section, we analyze the mean stability and steady state mean-square performance of the ILMS algorithm when communication channels between nodes are fading channels. Our analysis is based on the energy conservation approach of \cite{lopes07, sayed08}. 

In our analysis, we will use the deviation between an observed measurement and its prediction based on the current local estimate, which is defined as
\begin{align*}
\pmb{e}_k (i) &= \pmb{d}_k (i) - \pmb{u}_{k,i} \bw_{k - 1,i},
\end{align*}
and the weight error vector, which is the deviation between the local estimate $\bw_{k,i}$ and its true value $w^o$, given by
\begin{align*}
\widetilde{\bw}_{k,i} &= w^o- \bw_{k,i}.
\end{align*}

It must be noted that in mean and mean-square analysis of the ILMS algorithm with fading channels, we need to evaluate the first and second order moments of the weight error vector $\widetilde{\bw}_{k,i} $.

\subsection{Mean Stability Analysis}
Consider the update equation \eqref{films}. By subtracting $w^o$ from both sides of \eqref{films} and using the definition of $\widetilde{\bw}_{k,i}$ we obtain
\begin{align} \label{wer}
\widetilde{\bw}_{k,i} &= {\pmb{h}_k}(i)\widetilde{\bw} _{k - 1,i}  \nonumber\\
 &+ (1 - {\pmb{h}_k}(i)){w^o} - {\mu _k}{\pmb{h}_k}(i)\pmb{u}_{k,i}^ * {\pmb{u}_{k,i}}\widetilde{\bw} _{k - 1,i} \nonumber\\
 &- {\mu _k}\pmb{u}_{k,i}^ * {\pmb{v}_k}(i) - {\mu _k}(1 - {\pmb{h}_k}(i))\pmb{u}_{k,i}^ * {\pmb{u}_{k,i}}{w^o} - {\pmb{q}_{k,i}} \nonumber\\
 &+ {\mu _k}\pmb{u}_{k,i}^ * {\pmb{u}_{k,i}}{\pmb{q}_{k,i}}.
\end{align}
The recursion \eqref{wer} shows how the weight error vector $\widetilde{\bw}_{k,i}$ evolves over each update of the ILMS algorithm at time $i$. We use it to derive the required sufficient conditions for mean convergence of the ILMS algorithm with fading channels in the following result. 

\begin{prop}
Under Assumptions \ref{assumpt:model} and \ref{assumpt:fading}, the ILMS algorithm over fading channels, given by the update equation \eqref{films}, is stable in mean for any initial conditions if we have 
\begin{equation} \label{pro1}
\max \left\{0,\frac{m_k-1}{m_k \lambda_{\min}(R_{u,k})} \right\} < \mu_k < \frac{m_k+1}{m_k \lambda_{\max}(R_{u,k})}.
\end{equation}
\end{prop}
\begin{IEEEproof}
To derive the stability condition in the mean, we need to derive a condition on the step-size in order to guarantee convergence in the mean so that $\mathop {\lim }\limits_{i \to \infty } \mathbb{E}\left[ {{{\widetilde \bw}_{k,i}}} \right] = 0$. To this end we consider again \eqref{wer}. By taking expectation on both sides of \eqref{wer} and using Assumptions \ref{assumpt:fading} yields
\begin{equation} \label{exp}
\mathbb{E}\left[ \widetilde{\bw} _{k,i}\right] = m_k J_k \mathbb{E}\left[ \widetilde{\bw} _{k-1,i}\right]
 +(1-m_k) J_k w^o
\end{equation}
with
\begin{equation} \label{defj}
J_k \triangleq I-\mu_k R_{u,k}.
\end{equation}
By iterating \eqref{exp}, we obtain that the weight error vector evolves according to 
\begin{align} \label{expv}
\mathbb{E}\left[ \widetilde{\bw} _{k,{i}}\right] &= \left(\prod_{n=1}^{N} m_n J_n \right)  \mathbb{E}\left[ \widetilde{\bw} _{k,{i-1}}\right] \nonumber \\
&  +\sum_{n=1}^{N}\left((1-m_n) J_n  \prod_{\ell=n+1}^{N} m_{\ell} J_{\ell} \right) w^o. 
\end{align}
We can see from \eqref{expv} that convergence in mean for the ILMS algorithm with fading channels depends on the modes of the matrix  
\begin{equation} \label{mod}
\mathcal{M} \triangleq \prod_{n=1}^{N} m_n J_n.
\end{equation}
The necessary and sufficient condition required for convergence of \eqref{films} in the mean is that the matrix $\mathcal{M}$ be stable.  Equivalently, all eigenvalues of $\mathcal{M}$ must be inside the unit circle, i.e., 
\begin{align} \label{rho}
& \rho  \left(\mathcal{M} \right)<1, 
\end{align}
We already know that the spectral radius of any matrix $X$ we have $\rho \left(X \right)<\|X\|$ for any induced matrix
norm \cite{horn}. Thus we have
\begin{align}
 \rho  (\mathcal{M}) & \leq \|\mathcal{M} \|    \nonumber  \\
                     & \leq \|m_1 J_1 \|\|m_2 J_2 \| \cdots \|m_N J_N\|       \nonumber  \\
                     &= \rho ({m_1}{J_1})\rho ({m_2}{J_2}) \cdots \rho ({m_N}{J_N})   \nonumber 
\end{align}
Since every $\{J_k\}$ is a Hermitian matrix, its 2-induced norm agrees with its spectral radius, which explains the last equality. Thus, to guarantee that for any $k$ the constraint $\eqref{rho}$ is satisfied if for all $k=1,\ldots,N$, it is enough to have $\rho ({m_k}{J_k})\leq 1$ which can be stated in terms of the $\lambda$ of $R_{u,k}$ as
\begin{equation} \label{cons}
\left|m_k (1-\mu_k \lambda)\right|<1. 
\end{equation}
Therefore, if the step-size $\mu_k$ is chosen such that 
\begin{equation} \label{prof1}
 \max \left\{0,\frac{m_k-1}{m_k \lambda_{\min}(R_{u,k})} \right\} < \mu_k < \frac{m_k+1}{m_k \lambda_{\max}(R_{u,k})} \nonumber 
\end{equation}
convergence of $\mathbb{E}\left[ \widetilde{\bw} _{k,{i}}\right]$ in the mean is guaranteed, and the proof is complete.
\end{IEEEproof}

\begin{remark}
It is worth noting that for the traditional ILMS algorithm in \eqref{ilms}, where information is exchanged perfectly over ideal links (i.e., we have $m_k=1$ for all nodes $k$), the convergence condition \eqref{pro1} reduces to 
\begin{equation} \label{muid}
0<\mu_k<\frac{2}{\lambda_{\max} (R_{u,k})},
\end{equation}
which is the same as that derived in \cite{sayed13a}.
\end{remark}

\begin{remark} We can conclude from \eqref{pro1} that even in the presence of fading channels, it is possible to ensure convergence of the ILMS algorithm in the mean, by choosing $\mu_k$ sufficiently small for all nodes $k$. Furthermore, the range that the step size can be chosen depends only on the mean of the channel fading statistics. This implies that if the channel gain can be estimated so that the resulting deviation gives $m_k=1$, then the step size can be chosen to be the same as in the ideal link case. Furthermore, by comparing \eqref{pro1} with \eqref{muid}, we see that, in the presence of fading channels, the allowable range of step sizes for stability in mean can increase or decrease depending on the average channel gain.
\end{remark}

\begin{prop}\label{prop:biased}
Suppose that Assumptions \ref{assumpt:model} and \ref{assumpt:fading} hold, and that $\mu_k$ satisfies \eqref{pro1} for all $k=1,\ldots,N$. Then, for the ILMS algorithm over fading channels, we have
\begin{align*} 
& \mathop {\lim }\limits_{i \to \infty}  { \mathbb{E}\left[ \widetilde{\bw} _{k,{i}}\right]} \nonumber \\
& =  (I-\mathcal{M})^{-1} \sum_{n=1}^{N}\left((1-m_n) J_n  \prod_{\ell=n+1}^{N} m_{\ell} J_{\ell} \right) w^o. 
\end{align*}

\end{prop}
\begin{IEEEproof}
At steady state, we have $\lim_{i\to\infty}\E{\widetilde{\bw} _{k,{i}}}=\lim_{i\to\infty}\E{\widetilde{\bw} _{k,{i-1}}}$ for every node $k$. Taking the limit as $i\to\infty$ in \eqref{expv}, we obtain the desired result.
\end{IEEEproof}

\begin{remark} From the proof of Proposition \ref{prop:biased}, it follows that if $m_k=1$ for all nodes $k$, then the ILMS algorithm is asymptotically unbiased, which is consistent with the result for the mean convergence of the ILMS algorithm over ideal links or links with only additive noise \cite{azam11a,azam11b}. Furthermore, we see that if channel estimation achieves an error with average mean $m_k=1$ for all channels, then the ILMS algorithm will also produce an asymptotically unbiased estimate. On the other hand, if $m_k \ne 1$ for some of the nodes $k$, then it is possible for the ILMS algorithm to produce an estimate that is asymptotically biased. From Proposition \ref{prop:biased}, this bias is the same for all nodes in the network. This is reminiscent of the convergence result for the ideal link case, which states that all the estimators $\bw_{k,i}$ converge in the mean to $w^o$.
\end{remark}

\subsection{Steady State Mean-Square Performance Analysis}
We now consider the steady state mean-square performance of the ILMS algorithm over fading channels. We are interested to quantify the performance using the following metrics at every node $k$:
\begin{align} 
\eta_k & \triangleq   \mathop {\lim }\limits_{i \to \infty} \mathbb{E}\left[\|\widetilde{\bw}_{k-1,i}\|^2 \right]\     (\mathrm{MSD})         \label{msd} \\
\zeta _k  & \triangleq    \mathop {\lim }\limits_{i \to \infty} \mathbb{E}\left[ \| {\widetilde{\bw} _{k - 1,i} } \|_{R_{u,k} }^2 \right] \,({\rm{EMSE}} ) \label{emse}  \\
\xi_k & \triangleq   \mathop {\lim }\limits_{i \to \infty} \mathbb{E}\left[|\pmb{e}_{k}(i)|^2\right]= \zeta_k+\sigma_{v,k}^2\ (\mathrm{MSE})     \label{mse}
\end{align}
In general, to derive the above steady state performance metrics, we need to evaluate quantities of the form $\mathbb{E}\left[\| {\widetilde{\bw} _{k,i}}\|_{{\Sigma _k}}^2 \right]$ where $\Sigma_k$ is a positive semi-definite Hermitian matrix. To this end, we consider the weight vector update equation given by \eqref{wer}. Let
\begin{align}
C_{k,i} &=m_k J_k C_{k-1,i}+(1-m_k)J_k ,
\end{align}
where $C_{0,i} =C_{N,i-1}$, and  $C_{0,1}=I$. Note that the matrix $C_{k,i}$ is such that
\begin{equation}\label{eqn:Ck}
\mathbb{E}\left[\widetilde{\bw}_{k ,i}\right] = {C_{k ,i}}{w^o}.
\end{equation}
By equating the weighted norm of both sides of \eqref{wer}, taking expectations and using Assumptions \ref{assumpt:model} and \ref{assumpt:fading}, and \eqref{eqn:Ck}, we obtain the following recursive relationship:
\begin{align} \label{var}
 \mathbb{E}\left[\| {\widetilde{\bw} _{k,i}}\|_{\Sigma_k}^2 \right]
&= \mathbb{E}\left[ \| {\widetilde{\bw} _{k - 1,i}}\|_{{\Sigma'_k}}^2\right] + \mu _k^2\sigma _{v,k}^2 \mathbb{E}\left[\| {{\pmb{u}_{k,i}}}\|_{{\Sigma _k}}^2\right]  \nonumber\\
 & + \mathbb{E}\left[ \| {{\pmb{q}_{k,i}}}\|_{{G_k}}^2\right]+ \| {{w^o}}\|_{{T_k} + {H_{k,i}}}^2,
\end{align}
where 
\begin{align}
{G_k} &= {\Sigma _k} - {\mu _k} \mathbb{E}\left[{\Sigma _k}\pmb{u}_{k,i}^ * {\pmb{u}_{k,i}} + \pmb{u}_{k,i}^ * {\pmb{u}_{k,i}}{\Sigma _k}\right] \\ \nonumber
&   \hspace{3cm}  + \mu _k^2 \mathbb{E}\left[\| {{\pmb{u}_{k,i}}} \|_{{\Sigma _k}}^2\pmb{u}_{k,i}^ * {\pmb{u}_{k,i}}\right] \\
\Sigma'_k &= s_k {G_k}     \\
{T_k} &= (1-2m_k+s_k) {G_k}      \\
{H_{k,i}} &= (m_k-s_k) ({C_{k - 1,i}}{G_k} + {G_k}{C_{k - 1,i}}).  \label{hre} 
\end{align}
In order to compute all the moments that appear in the recursive equation \eqref{var} and to obtain closed-form expressions, we now make the following assumption regarding the regression vectors $\bu_k$, for all nodes $k=1,\ldots,N$.
\begin{assumption}\label{assumpt:Gaussian}
For each $k=1,\ldots,N$, the distribution of $\bu_k$ is a Gaussian distribution with
\begin{equation} \label{deco}
R_{u,k}= U_k \Lambda _k U_k^ *, 
\end{equation}
where $\Lambda _k $ is a diagonal matrix with diagonal elements being the eigenvalues of the correlation matrix $R_{u,k}$, and $U_k$ is unitary matrix.
\end{assumption}

Making use of Assumption \ref{assumpt:Gaussian}, we further define the following transformed quantities:
\begin{IEEEeqnarray*}{rCl's} 
{{\bar {\bw} }_{k,i}} &=& U_k^ * {\widetilde{\bw} _{k,i}},
& ${{\bar \Sigma }_k} = U_k^ * {\Sigma _k}{U_k}$, \\
{{\bar \Sigma '}_k} &=& U_k^ * {{\Sigma '}_k}{U_k}, 
& ${{\bar {\pmb{u}}}_{k,i}} = {\pmb{u}_{k,i}}{U_k}$, \\
{{\bar T}_k} &=& U_k^ * {T_k}{U_k},
& ${{\bar H}_{k,i}} = U_k^ * {H_{k,i}}{U_k}$, \\
{{\bar w}^o} &=& U_k^ * {w^o},
& ${{\bar C}_{k - 1, {i}}} = U_k^ * {C_{k - 1, {i}}}{U_k}$, \\
{{\bar Q}_k} &=& U_k^ * {Q_k}{U_k},
& {$D={{\bar w}^o}{{\bar w}^{o * }} = w^o w^{o*}$}.
\end{IEEEeqnarray*}
From the above definitions, equation \eqref{var} can now be rewritten in the following equivalent form
\begin{align} \label{tvar}
\mathbb{E} \left[\| {\bar {\bw} _{k,i}} \|_{{{\bar \Sigma }_k}}^2 \right] &= \mathbb{E} \left[\| {\bar {\bw} _{k - 1,i}} \|_{{{\bar \Sigma '}_k}}^2 \right]+ {\mu_k ^2}\sigma _{v,k}^2 \mathbb{E} \left[\| {{{\bar {\pmb{u}}}_{k,i}}} \|_{{{\bar \Sigma }_k}}^2 \right]  \nonumber\\
&   \hspace{1cm} + \mathbb{E}\left[\| {{{\bar {\pmb{q}}}_{k,i}}} \|_{{{\bar G}_k}}^2 \right] + \| {{{\bar w}^o}} \|_{{{\bar T}_k} + {{\bar H}_{k,i}}}^2
\end{align}
where in \eqref{tvar} we have 
\begin{align}
{{\bar G}_k} &= {{\bar \Sigma }_k} - {\mu _k} \mathbb{E}\left[{{\bar \Sigma }_k}\bar {\pmb{u}}_{k,i}^ * {{\bar {\pmb{u}}}_{k,i}} + \bar {\pmb{u}}_{k,i}^ * {{\bar {\pmb{u}}}_{k,i}} {{\bar \Sigma }_k}\right] \nonumber \\
& \hspace{2cm} + \mu _k^2 \mathbb{E} \| {{{\bar {\pmb{u}}}_{k,i}}} \|_{{{\bar \Sigma }_k}}^2\bar {\pmb{u}}_{k,i}^ * {{\bar {\pmb{u}}}_{k,i}} \nonumber 
\\
 &= {{\bar \Sigma }_k} - {\mu _k}({{\bar \Sigma }_k}{\Lambda _k} + {\Lambda _k}{{\bar \Sigma }_k})  \nonumber\\ 
& \hspace{2cm}  + \mu _k^2({\Lambda _k}{\rm{Tr[}}{{\bar \Sigma }_k}{\Lambda _k}]+  {\Lambda _k}{{\bar \Sigma }_k}{\Lambda _k})    \label{gbar}
\\
{{\bar \Sigma '}_k} &=  s_k {{\bar G}_k}   \label{sbar}
\\
{{\bar T}_k} &=  (1-2m_k+s_k) {{\bar G}_k}    \nonumber
\\
 {\bar H}_{k,i}&=    (m_k-s_k)  ({{\bar C}_{k - 1,i}}{{\bar G}_k} + {{\bar G}_k}{{\bar C}_{k - 1,i}}) \label{tbar} 
\end{align}
Further algebraic manipulations of \eqref{tvar} yields
\begin{align} \label{cvar}
\mathbb{E}\left[ \| {{{\bar {\bw}}_{k,i}}} \|_{{{\bar \Sigma }_k}}^2\right] &= \mathbb{E}\left[ \| {{{\bar {\bw} }_{k - 1,i}}} \|_{{{\bar \Sigma '}_k}}^2 \right]+ \mu _k^2\sigma _{v,k}^2{\rm{Tr[}}{\Lambda _k}{{\bar \Sigma }_k}] \nonumber\\
& + {\rm{Tr[}}{{\bar Q}_k}{{\bar G}_k}] + {\rm{Tr[}}D{{\bar T}_k}] + {\rm{Tr[}}D{{\bar H}_{k,i}}].
\end{align}
To derive \eqref{msd}-\eqref{mse}, we only need to consider the case where $\bar{\Sigma}{_k}$ is a diagonal matrix. In this case, matrix $\bar{\Sigma}^\prime{_k}$  is also a diagonal matrix. We let
\begin{equation} \label{dvec}
\bar{\sigma}_k   \triangleq {\rm{\diag}}\{ \bar{\Sigma}_k  \}, \quad \bar{\sigma}_k ^\prime   \triangleq {\rm{diag}}\{ \bar{\Sigma}_k  ^\prime  \},  \quad \lambda _k  \triangleq {\rm{diag}}\{ \Lambda _k \},
\end{equation}
and
\begin{equation} \label{fbar}
{{\bar F}_k} = I - \mu_k X_k +\mu_k^2 Y_k,
\end{equation}
with $X_k=2 {\Lambda _k}$ and $Y_k=\Lambda _k^2 + {\lambda _k}\lambda _k^T$. The $M \times M$ matrix $\bar F_k$ contains the statistics of data local to node $k$. We then have 
\begin{equation} \label{fvar}
\mathbb{E}\left[ \| {\bar {\bw} _{k,i}} \|_{{{\bar \sigma }_k}}^2 \right] = \mathbb{E}\left[ \| {\bar {\bw} _{k - 1,i}} \|_{{{\bar {\sigma} '}_k}}^2 \right]+ {{g}_{k,i}}{{\bar {\sigma}_k}},
\end{equation}
where $g_{k,i}$ and ${\bar {{\sigma}} '}_k$ are given  respectively by
\begin{align}
g_{k,i} &= \mu _k^2\sigma _{v,k}^2\lambda _k^T + \diag{\{ {{\bar Q}_k}\} ^T}{{\bar F}_k} \nonumber \\
&  \hspace{.1cm}+ (1-2m_k+s_k)  \diag{\{ D\} ^T}{{\bar F}_k} \nonumber \\ 
&  \hspace{.1cm} + 2(m_k-s_k) \diag{\{ D\} ^T}{{\bar C}_{k - 1,i}}{{\bar F}_k}, \nonumber \\
{\bar {{\sigma}} '}_k &=  s_k {{\bar F}_k}{{\bar {{\sigma}} }_k}   \label{sfor}.
\end{align}

We next use \eqref{fvar} to derive conditions that guarantee convergence in the mean-square sense for the ILMS algorithm with fading channels. 
\begin{prop}\label{prop:MSconv}
Under assumptions \ref{assumpt:model}, \ref{assumpt:fading} and \ref{assumpt:Gaussian}, the ILMS algorithm over fading channels converges in the mean-square sense if the step sizes $\mu_k$ are chosen to be sufficiently small so that the 
\begin{align}\label{msconvcond}
s_k \rho(\bar{{F}}_k)  &< 1.
\end{align}
\end{prop}
\begin{IEEEproof}
From \eqref{fvar} and using the results in Section 6.9 of \cite{sayed13a}, the ILMS algorithm with fading channels is stable in the mean-square sense if matrix $\bar{{F}}'_k \triangleq s_k\bar F_k$ is a stable matrix. Therefore, if we select the step size $\mu_k$ to be sufficiently small so that $\rho(\bar{{F}}'_k)<1$, then stability in the mean-square sense is guaranteed. The proof is now complete.
\end{IEEEproof}
\begin{remark}
Suppose that $s_k = s$ for all nodes $k$, and the step sizes $\mu_k$ are fixed. Then, if $s$ is sufficiently large, the left hand side of \eqref{fvar} diverges and we no longer have mean-square stability. This shows that deteriorating fading conditions have detrimental impact on the ILMS algorithm, and care should be taken to adjust the step sizes according to Proposition \ref{prop:MSconv}.  
\end{remark}

Assuming that step sizes are chosen sufficiently small, and by letting $i\to\infty$, the recursive equation \eqref{fvar} at steady-state gives
\begin{equation} \label{fvar_steady}
\mathbb{E}\left[ \| {\bar {\bw} _{k,\infty}} \|_{{{\bar \sigma }_k}}^2 \right] = \mathbb{E}\left[ \| {\bar {\bw} _{k - 1,\infty}} \|_{{{\bar {\sigma} '}_k}}^2 \right]+ {{g}_k}{{\bar {\sigma}_k}},
\end{equation}
where 
\begin{align}
g_{k} &= \mu _k^2\sigma _{v,k}^2\lambda _k^T + \diag{\{ {{\bar Q}_k}\} ^T}{{\bar F}_k} \nonumber \\
&  \hspace{.1cm}+ (1-2m_k+s_k)  \diag{\{ D\} ^T}{{\bar F}_k} \nonumber \\ 
&  \hspace{.1cm} + 2(m_k-s_k) \diag{\{ D\} ^T}{{\bar C}_{k - 1,\infty}}{{\bar F}_k}, \label{gfor}
\end{align}
and using \eqref{exp} and \eqref{eqn:Ck}, ${{\bar C}_{k - 1,\infty}}$ in \eqref{gfor} is given by
\begin{align} \label{Ckinf}
{{\bar C}_{k - 1,\infty}} &=U_k^* \Bigg( (I-\mathcal{M})^{-1} \times \nonumber \\
&  \hspace{1.2cm} \sum_{n=1}^{N} \bigg((1-m_n) J_n  \prod_{\ell=n+1}^{N} m_{\ell} J_{\ell} \bigg) \Bigg) U_k.
\end{align}

We observe that \eqref{fvar_steady} shows how $\mathbb{E}\left[ \| {\bar {\bw} _{k,\infty}} \|_{{{\bar \sigma }_k}}^2 \right]$ evolves through the network, which in its current form makes it difficult to derive the desired metrics \eqref{msd}-\eqref{mse} directly. In fact, we have to find a recursive equation that reveals how $\mathbb{E}\left[ \| {\bar {\bw} _{k,i}} \|_{{{\bar \sigma }_k}}^2 \right]$ evolves in time. By iterating \eqref{fvar}, and using $\bw_{0,i+1}= \bw _{N,i}$, we can obtain a set of $N$ coupled equations. With suitable manipulation of these equations, along with proper selections of ${{\bar {\sigma}}_k}$, it is possible to solve the resulting equalities to derive the desired metrics. Following the argument given in \cite{lopes07}, we can derive the required metrics in a similar way. For completeness, we provide an outline of the proof in Appendix \ref{appen:a}.

\begin{prop}\label{prop:fmsd}
Under Assumptions \ref{assumpt:model}, \ref{assumpt:fading} and \ref{assumpt:Gaussian}, the steady state performance of the ILMS algorithm over fading channels for each node $k$ in the mean-square sense is given by the following expressions: 
\begin{align}
\eta _k  &= a_k (I - \Pi _{k,1} )^{ - 1} \mathds{1}  \label{fmsd} \\ 
 \zeta _k  &= a_k (I - \Pi _{k,1} )^{ - 1} \lambda _k   \label{femse} \\
 \xi _k  &= \zeta _k  + \sigma _{v,k}^2   \label{fmse}
\end{align}
where 
\begin{align} \label{pfor}
\Pi _{k,l}  & \triangleq \bigg(\prod_{k=1}^{N} s_k\bigg) \bigg(\bar{F} _{k + l - 1}  \bar{F} _{k + l}  \cdots  \bar{F} _N   \bar{F} _1  \cdots \bar{F} _{k - 1}\bigg), \\
a_k  & \triangleq g_k \Pi _{k,2}  + g_{k + 1} \Pi _{k,3}  +  \ldots  + g_{k - 2} \Pi _{k,N}  + g_{k - 1}, \label{afor}
\end{align}
where $l = 1, \cdots ,N$ and all the subscripts are in $\bmod \;N$.
\end{prop}

\begin{remark} Note that $g'_k = \mu _k^2\sigma _{v,k}^2\lambda _k^T$ is the equivalent expression for \eqref{gfor} for the original ILMS algorithm over ideal communication links (see equation (55) in \cite{lopes07}). By comparing \eqref{gfor} with equation (55) in \cite{lopes07}, we can see that we can model the effect of fading channels as additional terms to $g'_k$ with
\begin{align}
g_k &= g'_k+ \diag{\{ {{\bar Q}_k}\} ^T}{{\bar F}_k} + (1-2m_k+s_k) \diag{\{ D\} ^T}{{\bar F}_k} \nonumber\\ 
 & \hspace{.5cm} + 2 (m_k-s_k) \diag{\{ D\} ^T}{{\bar C}_{k - 1,\infty}}{{\bar F}_k}.  \label{extra}
\end{align}
The first additional term on the right hand side of \eqref{extra} is due to the additive noise of the fading channel. The rest of the additional terms are due to statistics of the fading channel.
\end{remark}

\subsection{Dependence of MSD, EMSE and MSE on Channel Gain Variances}

In this subsection, we investigate the dependence of the MSD, EMSE and MSE on the channel gain variance or second order moment by making some assumptions in order to simplify the analysis. We show that the MSD, EMSE and MSE are non-decreasing functions of $s_k$. To do this, we adopt the same Gaussian model used in the previous subsection, with the further assumptions that
\begin{enumerate}
  \item $m_k=1,\ \ \  R_{u,k}=\lambda I,\ \ \ Q_k=\sigma_{c,k}^2 I$, and
	\item $\mu_k=\mu$ and $\mu$ is sufficiently small so that \eqref{msconvcond} holds, and ${{\bar F}_k}$ can be approximated as
\begin{align*} 
{{\bar F}_k} & \approx I - \mu X=(1-2 \mu \lambda) I. 
\end{align*}
\end{enumerate}
Since ${{\bar F}_k}$ is now a diagonal matrix, the matrix $\Pi_{k,\ell} \approx {\bar F}_1 {\bar F}_2\cdots {\bar F}_N$ is also diagonal, and can be approximated as 
\begin{align} \label{Papprox}
\Pi_{k,\ell} &  \approx  s_{p}(1-2\mu \lambda)^{ N } I, 
\end{align}
with $s_p \triangleq \prod_{k=1}^{N} s_k$. Using \eqref{Papprox} we have
\begin{align}\label{IminPi}
I-\Pi_{k,\ell} &  \approx \left(1-s_{p}(1-2\mu \lambda)^{ N }\right) I  
\end{align}
Similarly, we can also obtain approximations for $g_k$ and $a_k$ as follows: 
\begin{align}
g_k &  \approx (\mu^2 \sigma_{v,k}^2 \lambda) \mathds{1}^T +(\sigma_{c,k}^2 (1-2\mu \lambda)) \mathds{1}^T \nonumber \\
& \hspace{2cm} + s_p(s_k-1)(1-2 \mu \lambda)\diag{\{D\}}^T  \nonumber \\
  &  \approx \left(\mu^2 \sigma_{v,k}^2 \lambda +\sigma_{c,k}^2 (1-2\mu \lambda) )\right) \mathds{1}^T     \nonumber \\
& \hspace{2cm} + s_p(s_k-1)(1-2 \mu \lambda)\diag{\{D\}}^T   \nonumber \\
a_k & \approx s_p(1-2 \mu \lambda)^{N} \Bigg( \sum_{n=1}^{N}g_n-g_{k-1}\Bigg)+g_{k-1}    \nonumber \\
    & \approx  \left(s_p(1-2 \mu \lambda)^{N}\right) \sum_{n=1}^{N}g_n + (s_p(1-2 \mu \lambda)^{N})g_{k-1}. \label{aprox}
\end{align}
Let $\hat{g}_k \triangleq g_k \mathds{1}$. Since $\diag{\{D\}}= [|w^o(1)|^2, \cdots,|w^o(M)|^2]^T$, we have
\begin{align*}
\hat{g}_k &=\mu^2 \sigma_{v,k}^2 \lambda +\sigma_{c,k}^2 (1-2\mu \lambda)  \nonumber \\
& \hspace{2cm} + s_p(s_k-1)(1-2 \mu \lambda)\|w^o\|^2.
\end{align*}
Replacing \eqref{IminPi} and \eqref{aprox} in \eqref{fmsd} we finally have
\begin{align} \label{msdaprox}
\eta_k & \approx \left(\frac{1}{1-s_p(1-2 \mu \lambda)^{N}}-1 \right)
\sum_{n=1}^{N} \hat{g}_n+\hat{g}_{k-1}.
\end{align}
Similarly, we obtain 
\begin{align} \label{emseaprox}
\zeta_k & \approx \lambda \left(\frac{1}{1-s_p(1-2 \mu \lambda)^{N}}-1 \right)
\sum_{n=1}^{N} \hat{g}_n+\lambda \hat{g}_{k-1},
\end{align}
and the approximation for the MSE $\xi_k$ follows from \eqref{fmse}.
Since $\hat{g}_n$ is a non-decreasing and non-negative function of $s_k$ for all $k$, from \eqref{msdaprox} and \eqref{emseaprox}, we see that the MSD, EMSE and MSE are all non-decreasing functions of $s_k$ (or equivalently the channel gain variances since we have assumed that the channel gain mean is 1).

Suppose further that $s_k=s\geq 1$ for all $k=1,\ldots,N$, and the step size $\mu$ is chosen so that $s(1-2 \mu \lambda)$ is a constant in $(0,1)$ for all values of $s$. In this case, from \eqref{msdaprox} and \eqref{emseaprox}, we obtain $\eta_k = \mathcal{O}(N s^{N+1})$ and $\zeta_k = \mathcal{O}(N s^{N+1})$.\footnote{A non-negative function $f(x)$ is $\mathcal{O}(g(x))$ if there exists $a > 0$ and $x_0$ such that for all $x \geq x_0$, $f(x) \leq a g(x)$.} This means that under the dB scale, the MSD, EMSE and MSE are approximately linear with respect to the channel gain second order moment. Simulations in Section~\ref{sect:Simulation} are presented to verify our conclusions.

\section{Simulation Results}\label{sect:Simulation}
\begin{figure}[t]
\centering 
\includegraphics [width=7cm]{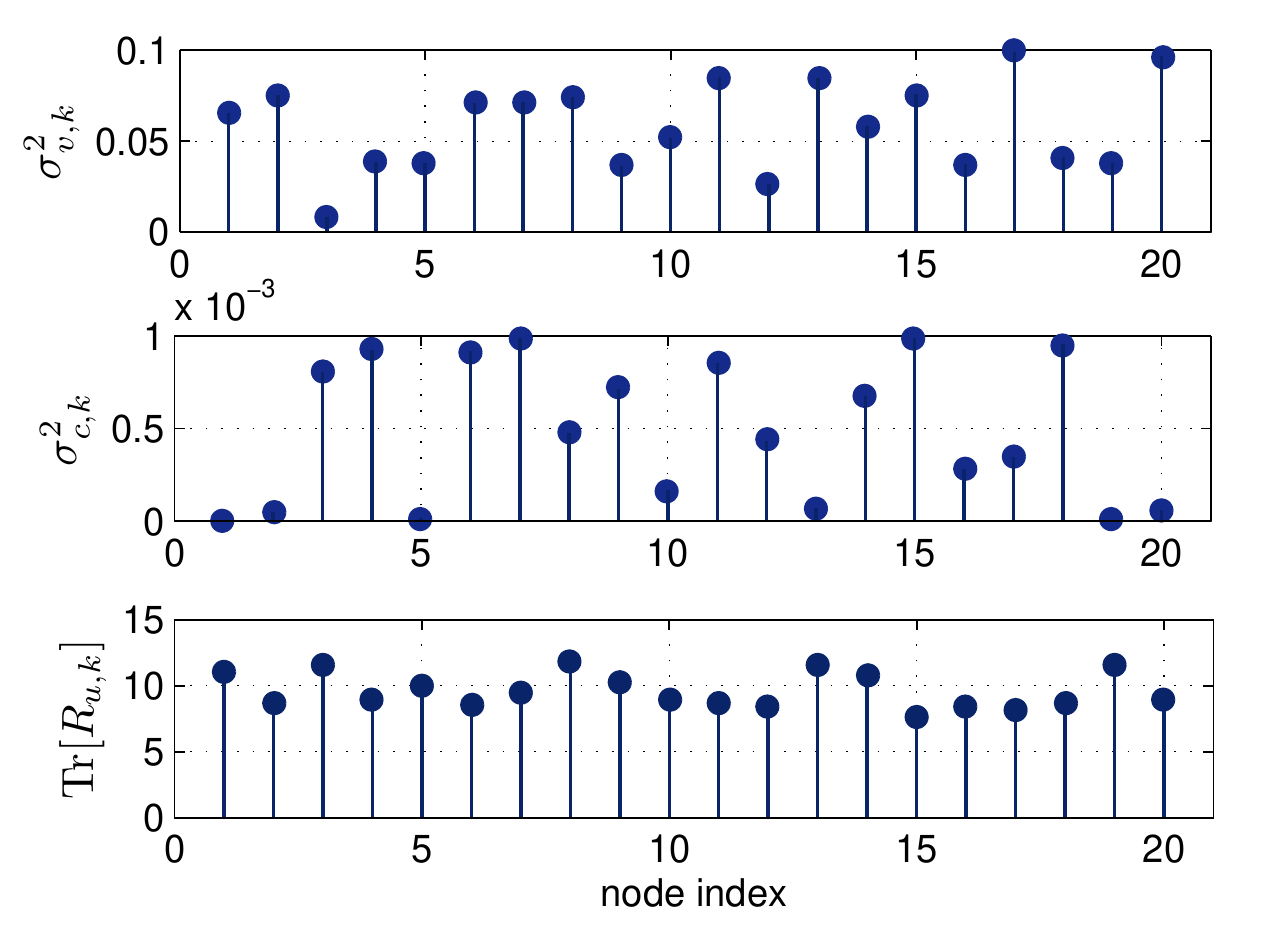} 
\centering \caption{Node profile and channel noise information: $\sigma_{v,k}^2$ (up), $\sigma_{c,k}^2$ (middle) and $\mathrm{Tr}(R_{u,k})$ (down).}
\label{fig:prof}
\end{figure}

To verify our theoretical performance analysis, we present some simulation results and compare them with the results in Section \ref{sect:Performance}. We assume a network composed of $N=20$ nodes,  where the nodes are connected via a ring topology as in the ILMS algorithm. The regressors $\bu_{k,i}$ are generated as independent realizations of a Gaussian distribution with a covariance matrix $R_{u,k}$ whose eigenvalue spread is 4. The measurement data $\bd_k(i)$ at each node $k$ is generated by using the data model \eqref{dmodel} where the parameter $w^o$ is chosen to be $[1~1~1~1]^T/2$, and the observation noise $\bv_k(i)$ is drawn from a Gaussian distribution with variance $\sigma_{v,k}^2$ as shown in Figure \ref{fig:prof}.  The additive channel noises are generated from Gaussian distributions with covariance matrix $Q_k=\sigma_{c,k}^2 I$, for $k=1,\ldots,20$. The values of $\sigma_{c,k}^2$ are shown in Figure \ref{fig:prof}. We generate the channel gains $\bh_k(i)$ using a Rayleigh distribution with $m_k=\sqrt{2}/2$ for all values of $k$. To obtain the steady-state values of MSD, EMSE and MSE, we run the ILMS algorithm with 2000 iterations and average the last 200 samples. Finally, each steady-state value is obtained by averaging over 100 independent runs. 

\begin{figure}[t]
\centering 
\includegraphics [width=7cm]{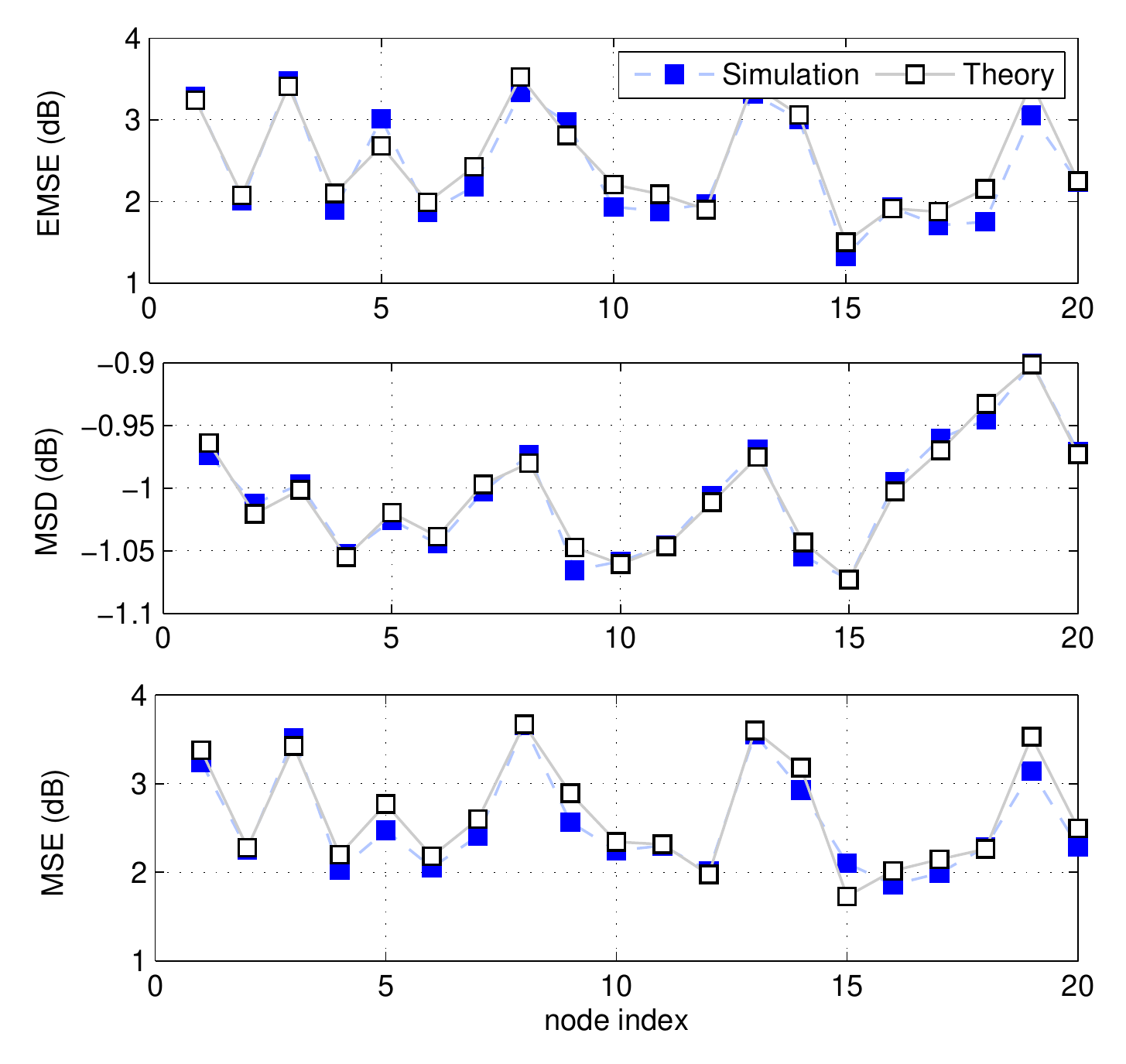} 
\centering \caption{Steady-state curves versus for each individual node $k$, $\mu=0.02$.}
\label{fig:std}
\end{figure}

\begin{figure}[t]
\centering 
\includegraphics [width=7cm]{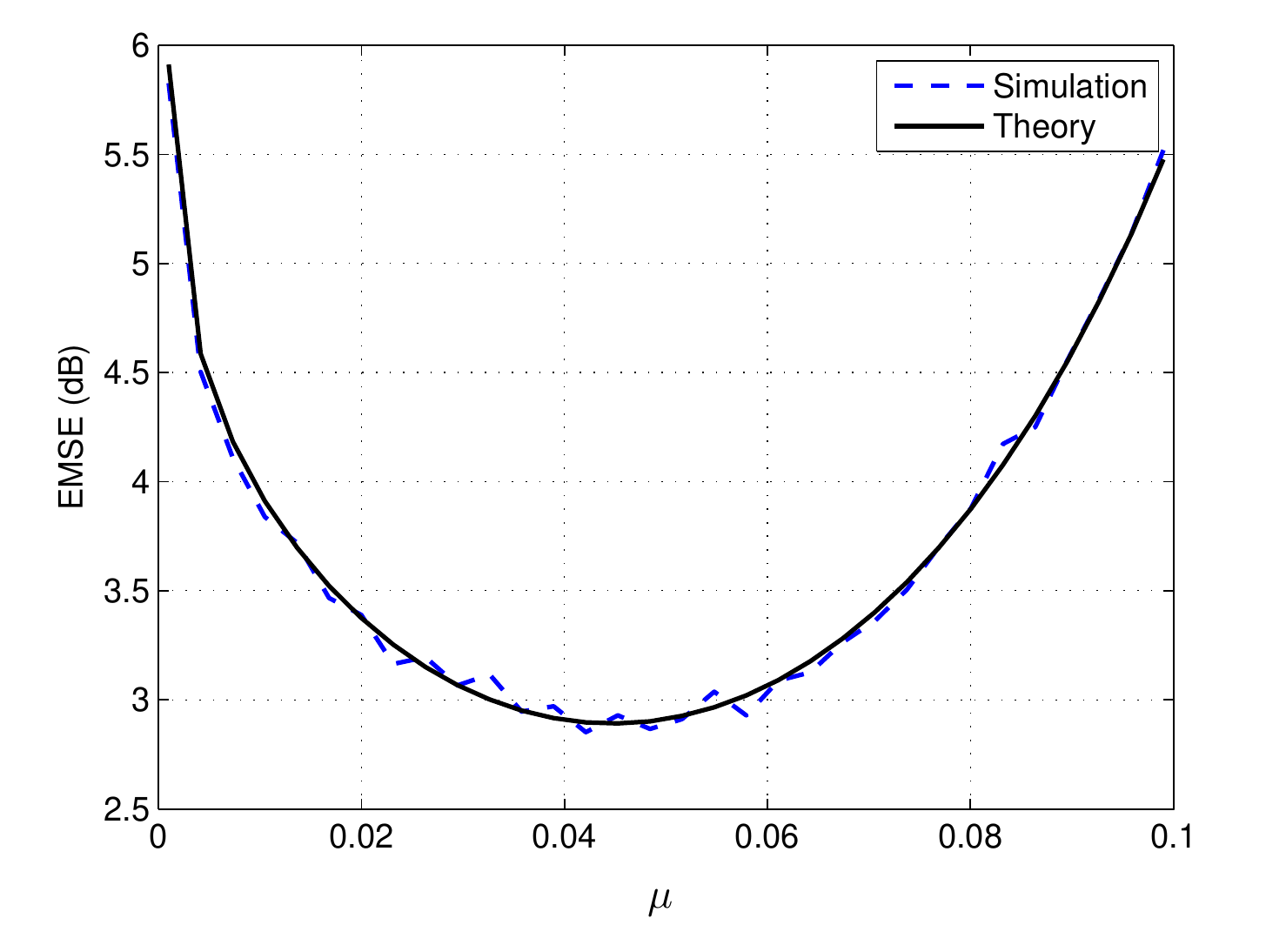} 
\centering \caption{Steady-state EMSE versus $\mu$ for node $k=1$.}
\label{fig:vmu}
\end{figure}

In Figure \ref{fig:std}, we show the steady-state performance metrics MSD, EMSE and MSE as functions of the node index $k$ when the step size $\mu=0.02$. We can see that the simulated results closely match the theoretical results. In Figure \ref{fig:vmu}, we have plotted the steady-state EMSE at node $k=1$ versus step size $\mu$.  We see that in contrast to the ideal link case, the curve is no longer a monotonically increasing function of the step-size $\mu$. 

\begin{figure}[t]
\centering 
\includegraphics [width=7cm]{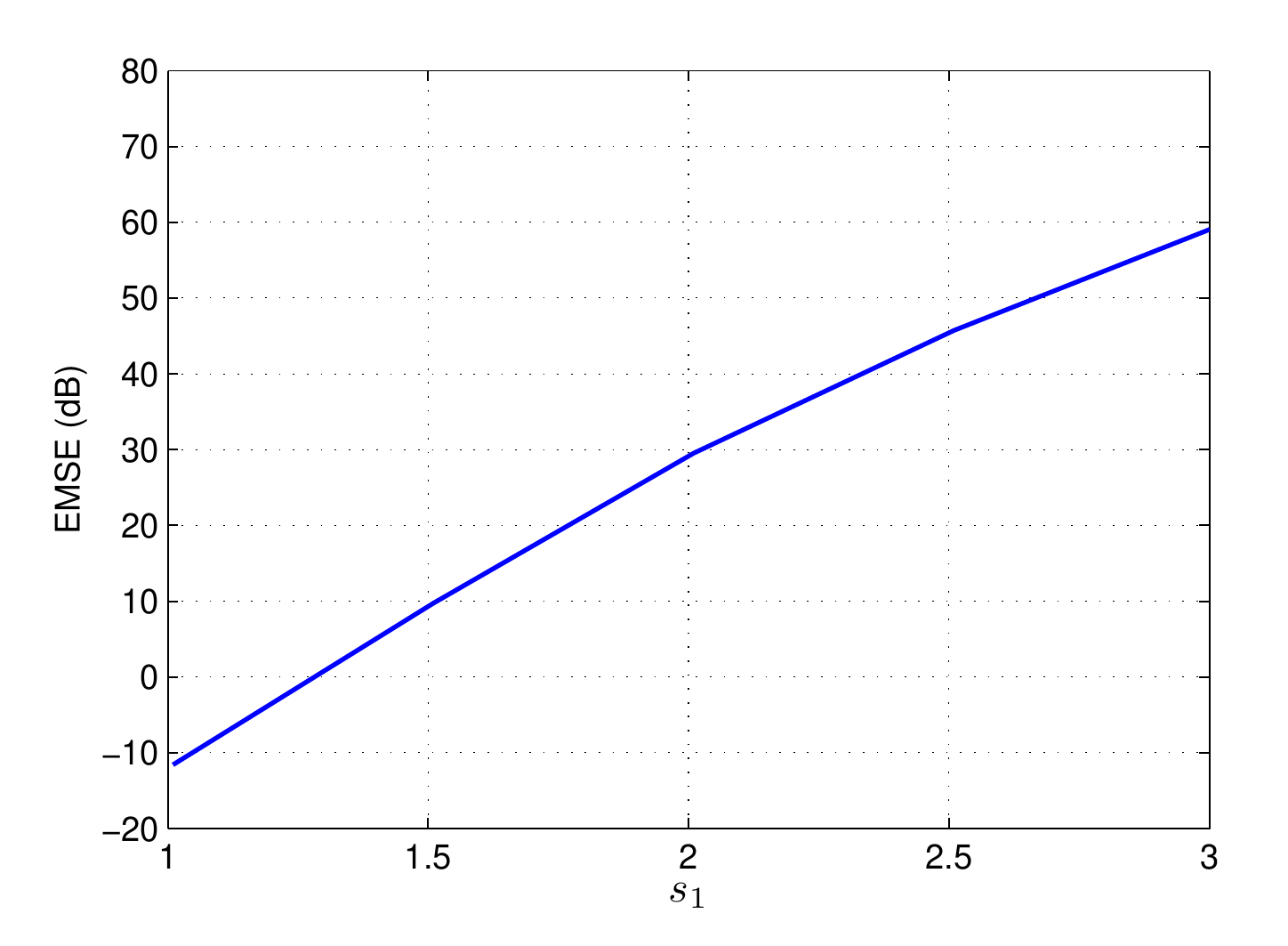} 
\centering \caption{Steady-state EMSE versus channel gain second order moment $s$ for node $k=1$.}
\label{fig:var}
\end{figure}

To show how the channel statistics affect the estimation performance, we simulate the case where all channels have average gain $1$. We further let $s_k = s$ for all $k=1,\ldots,20$, and evaluate the EMSE performance when $s$ increases. We plot the steady-state EMSE at node $k=1$ versus varying values of $s$ in Figure \ref{fig:var}. The step size is chosen as $\mu=0.02$. It can be seen that as $s$ increases, we get worse performance. On the other hand, as $s\rightarrow 0$, the performance of the ILMS algorithm over fading channels tend to the steady-state performance of the ILMS algorithm with additive noise links. Similar behavior is also seen for other steady-state metrics, i.e. MSD and MSE. This simulation also verifies our prediction of the linear relationship between EMSE and $s$ in dB.

\begin{figure}[!htb]
\centering 
\includegraphics [width=7cm]{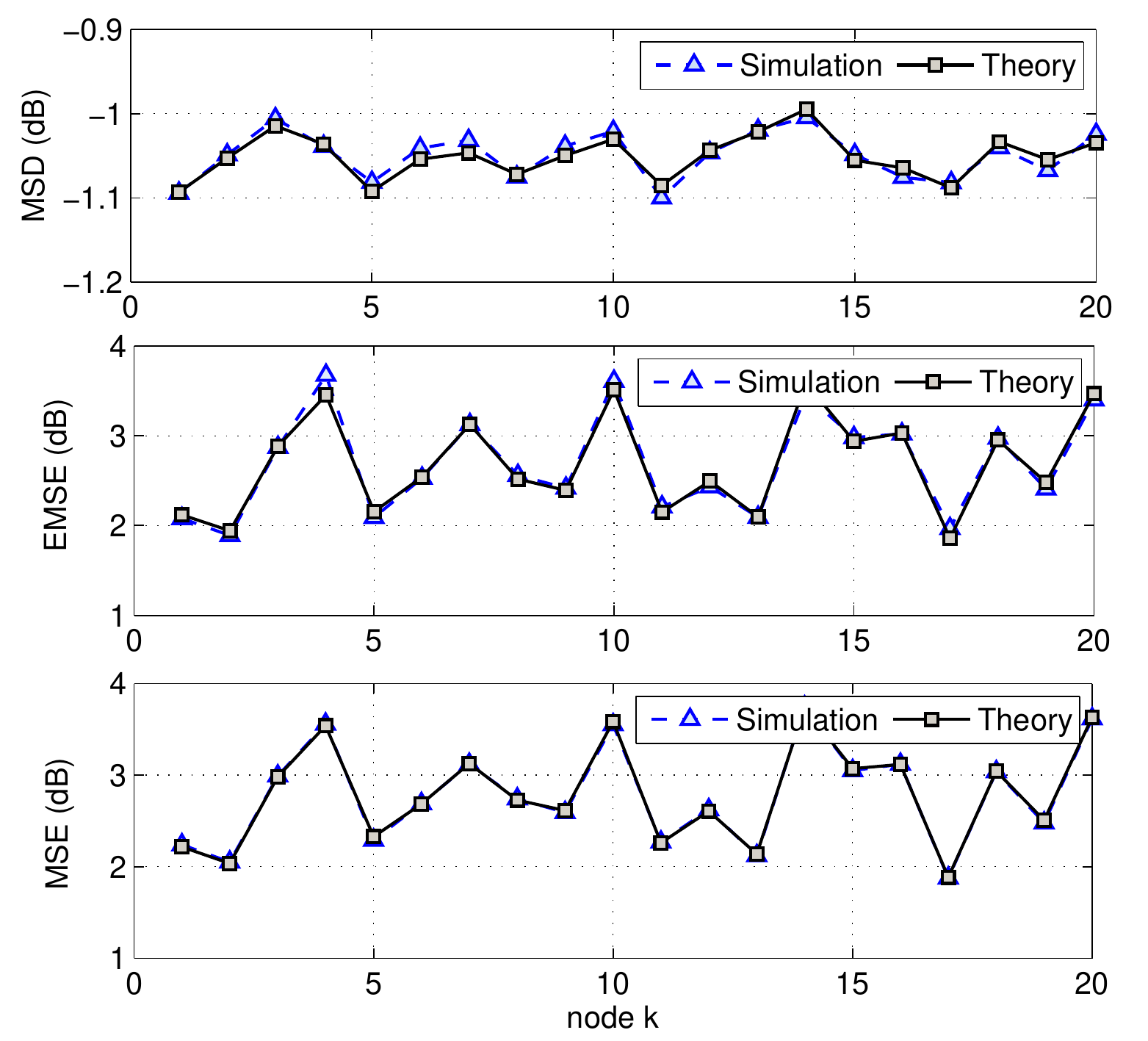} 
\centering \caption{Steady-state curves versus for each individual node $k$, $\mu=0.02$.}
\label{fig:stdsr}
\end{figure}

\begin{figure}[!htb]
\centering 
\includegraphics [width=7cm]{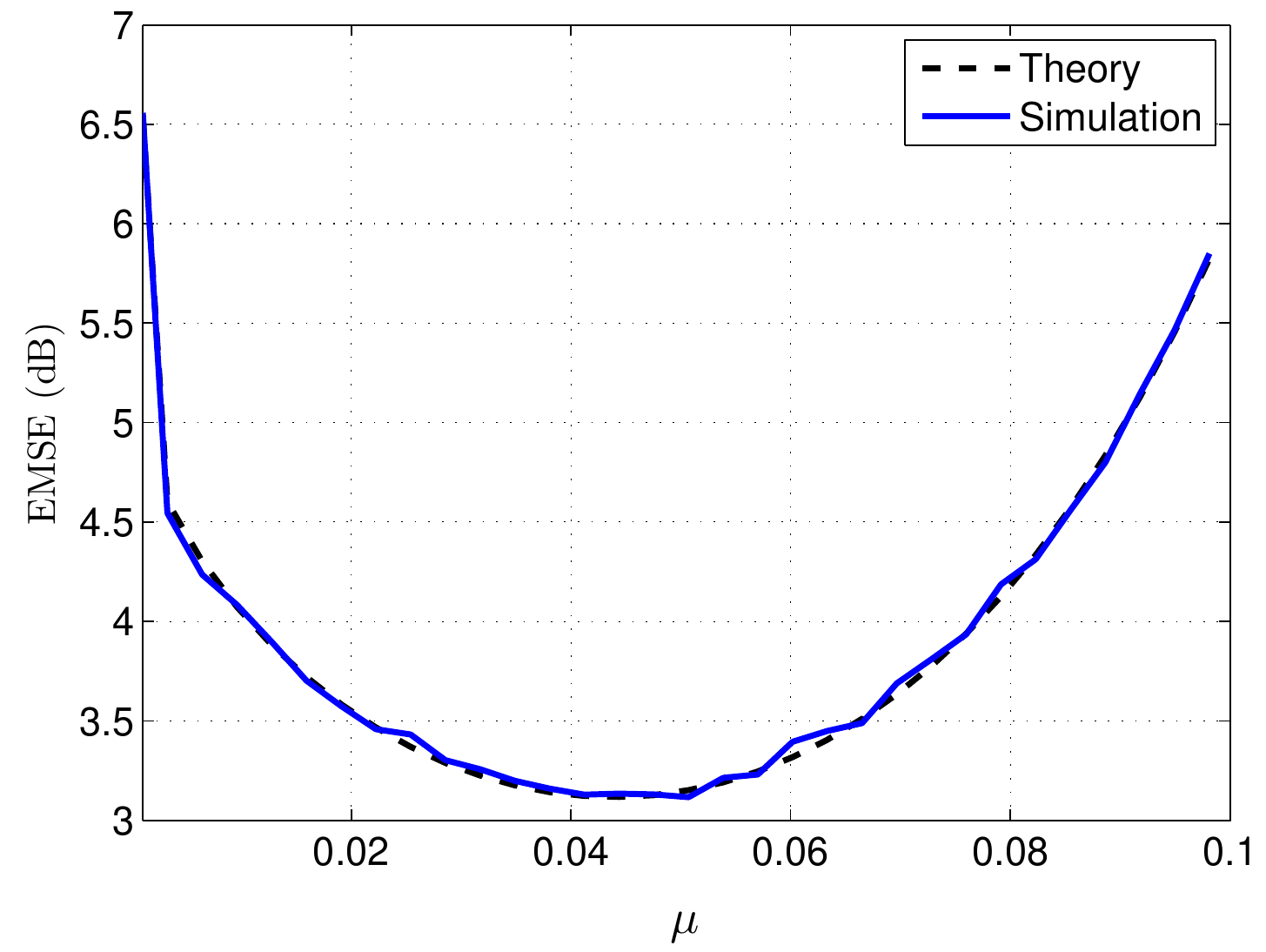} 
\centering \caption{Steady-state EMSE versus $\mu$ for node $k=1$. We have considered regression vectors with shift structure.}
\label{fig:vmushift}
\end{figure}

Although we assume that the regression vectors $\bu_{k,i}$ are independent in this paper, our simulations indicate that the theoretical results also hold approximately for regression vectors with a shift structure \cite{lopes07}. To show this, we suppose that at each node $k$, and for all times $i$, the regression vector $\bu_{k,i}$ can be expressed as
\begin{align*}
\bu_{k,i}&=[u_k(i), u_k(i-1),\cdots,u_k(i-M+1)],
\end{align*}
where $u_k(i)$ is generated according to the following recursion 
\begin{align} \label{shreg}
u_k(i)&=\alpha_k u_k(i-1)+\beta_k\tau_k(i),
\end{align}
which represents a first-order autoregressive (AR) process where with a pole at $\alpha_k$. In \eqref{shreg}, $\tau_k$ denotes a white, zero-mean, Gaussian
random sequence with unit variance, while we choose $\alpha_k\in(0,0.5]$ randomly and $\beta_k=\sqrt{\sigma_{u,k}^2(1-\alpha_k^2)}$. In Figure \ref{fig:stdsr}, we show the steady-state performance metrics MSD, EMSE and MSE as functions of the node index $k$. We can see that for this case the simulated results have good match with the theoretical derivations. We also have plotted the steady-state EMSE at node $k = 1$ versus step size $\mu$ in Fig. \ref{fig:vmushift}.

\section{Conclusion}\label{sect:Conclusion}
In this paper, we have investigated the steady state performance of the ILMS algorithm when the links between nodes are fading channels, and we do not have perfect channel state information. Our analysis reveals some interesting behaviors, including (i) in the presence of fading channels, the ILMS algorithm is asymptotically biased, (ii) a dynamic range for the step-sizes for mean stability that depends only on the mean channel gain can be derived and (iii) mean-square stability depends on the channel gain variances, with the MSD, EMSE and MSE being non-decreasing functions of these variances. We present simulation results to verify our theoretical analysis. Simulation results also show that the EMSE performance degrades as the fading channel gain variance increases.

In this paper, we focus on analyzing the effect of fading channels on incremental updates for the LMS filter. We believe that most of the insights obtained from this analysis are applicable to the more general class of diffusion LMS algorithms, whose performances under fading channels deserve further research. 

\appendices 
\section{Outline Proof of Proposition \ref{prop:fmsd}}\label{appen:a}
In this appendix, we provide an outline proof of Proposition \ref{prop:fmsd} for completeness. The proof steps are the same as those given in \cite{lopes07}, except with additional $m_k$ and $s_k$ terms. 

As we mentioned before, \eqref{fvar_steady} is a coupled equation since it involves both $\mathbb{E}\left[ \| {\bar {\bw} _{k,i}} \|_{{{\bar \sigma }_k}}^2 \right]$ and $\mathbb{E}\left[ \| {\bar {\bw} _{k,i}} \|_{{{\bar \sigma }_{k-1}}}^2 \right]$, i.e., information from two spatial nodes. However, we need to find a recursive equation that reveals how $\mathbb{E}\left[ \| {\bar {\bw} _{k,i}} \|_{{{\bar \sigma }_k}}^2 \right]$ evolves in time. We can exploit the ring (incremental) topology together with the weighting matrices to resolve this difficulty. To this end, let $\bp_k \triangleq {\bar {\bw} _{k,\infty}}$. By iterating \eqref{fvar_steady} and using \eqref{sfor},  we obtain the following coupled equalities 

\begin{align} 
\mathbb{E}\left[ \| {\bp_{1}} \|_{{{\bar \sigma }_1}}^2 \right] &= \mathbb{E}\left[ \| {\bp_{N}} \|_{s_1 {{\bar F}_1}{{\bar {{\sigma}} }_1}}^2 \right]+ {{g}_1}{{\bar {\sigma}_1}}, \nonumber \\
\mathbb{E}\left[ \| {\bp_{2}} \|_{{{\bar \sigma }_2}}^2 \right] &= \mathbb{E}\left[ \| {\bp_{1}} \|_{s_2 {{\bar F}_2}{{\bar {{\sigma}} }_2}}^2 \right]+ {{g}_2}{{\bar {\sigma}_2}} \nonumber \\
\vdots  \nonumber \\
\mathbb{E}\left[ \| {\bp_{k-2}} \|_{{{\bar \sigma }_{k-2}}}^2 \right] &= \mathbb{E}\left[ \| {\bp_{k - 3}} \|_{s_{k-2} {{\bar F}_{k-2}}{{\bar {{\sigma}} }_{k-2}}}^2 \right]+ {{g}_{k-2}}{{\bar {\sigma}_{k-2}}}  \label{coeq1}  \\
\mathbb{E}\left[ \| {\bp_{k-1}} \|_{{{\bar \sigma }_{k-1}}}^2 \right] &= \mathbb{E}\left[ \| {\bp_{k - 2}} \|_{s_{k-1} {{\bar F}_{k-1}}{{\bar {{\sigma}} }_{k-1}}}^2 \right]+ {{g}_{k-1}}{{\bar {\sigma}_{k-1}}} \label{coeq2} \\
\vdots  \nonumber \\
\mathbb{E}\left[ \| {\bp_{N}} \|_{{{\bar \sigma }_N}}^2 \right] &= \mathbb{E}\left[ \| {\bp_{N - 1}} \|_{s_N {{\bar F}_N}{{\bar {{\sigma}} }_N}}^2 \right]+ {{g}_N}{{\bar {\sigma}_N}}
\end{align}
We can see from \eqref{coeq2} that $\mathbb{E}\left[ \| {\bp_{k-1}} \|_{{{\bar \sigma }_{k-1}}}^2 \right]$ in term of $\mathbb{E}\left[ \| {\bp_{k - 2}} \|_{s_{k-1} {{\bar F}_{k-1}}{{\bar {{\sigma}} }_{k-1}}}^2 \right]$. Now, if we choose the weighting vector ${{\bar \sigma }_{k-2}}=s_{k-1} {{\bar F}_{k-1}}{{\bar {{\sigma}} }_{k-1}}$ in \eqref{coeq1} we get

\begin{align} \label{coeq3}
\mathbb{E}\left[ \| {\bp_{k-2}} \|_{s_{k-1} {{\bar F}_{k-1}}{{\bar {{\sigma}} }_{k-1}}}^2 \right] &= \mathbb{E}\left[ \| {\bp_{k - 3}} \|_{s_{k-1} s_{k-2}{{\bar F}_{k-2}} {{\bar F}_{k-1}}{{\bar {{\sigma}} }_{k-1}} }^2 \right] \nonumber \\
& \hspace{1cm}+ s_{k-1} {{g}_{k-2}} {{\bar F}_{k-1}}{{\bar {{\sigma}} }_{k-1}}
\end{align}
Replacing \eqref{coeq3} in \eqref{coeq2} yields 
\begin{align} \label{coeq4}
\mathbb{E}\left[ \| {\bp_{k-1}} \|_{{{\bar \sigma }_{k-1}}}^2 \right] &= \mathbb{E}\left[ \| {\bp_{k - 3}} \|_{s_{k-1} s_{k-2}{{\bar F}_{k-2}} {{\bar F}_{k-1}}{{\bar {{\sigma}} }_{k-1}} }^2 \right] \nonumber \\
& \hspace{1cm}+ s_{k-1} {{g}_{k-2}} {{\bar F}_{k-1}}{{\bar {{\sigma}} }_{k-1}} \nonumber \\
& \hspace{1cm} + {{g}_{k-1}}{{\bar {\sigma}_{k-1}}}
\end{align}
Iterating in this manner, we can obtain an expression for $\mathbb{E}\left[ \| {\bp_{k-1}} \|_{{{\bar \sigma }_{k-1}}}^2 \right]$ as follows
\begin{align} \label{coeqfin}
 E \left[ \|\bp _{k - 1} \|_{\sigma _{k - 1} }^2 \right]  &= E\left[\bp _{k - 1} \|_{F'_k  \ldots F'_N F'_1  \ldots F'_{k - 1} \sigma _{k - 1} }^2\right]   \nonumber \\ 
 &  + g_k F'_{k + 1}  \ldots F'_N F_1  \ldots F'_{k - 1} \sigma _{k - 1}  \nonumber  \\
 &  + g_{k + 1} F'_{k + 2}  \ldots F'_N F'_1  \ldots F'_{k - 1} \sigma _{k - 1} \nonumber  \\
 &  +\cdots + g_{k - 2} F'_{k - 1} \sigma _{k - 1}  + g_{k - 1} \sigma _{k - 1}
\end{align}

Using \eqref{pfor} and \eqref{afor} we can represent \eqref{coeqfin} in the following form 
\begin{align} \label{gmetr}
E\left[ \left\| {\bp_{k - 1} } \right\|_{(I - \Pi _{k,1} )\sigma _{k - 1} }^2\right]  
&= a_k \sigma _{k - 1}.
\end{align}
Finally, we use equation \eqref{gmetr} to derive the required metrics at node $k$. In fact, since we are free to select the weight vector $\sigma_{k-1}$, choosing $\sigma_{k-1}=(I - \Pi _{k,1} )^{- 1} {\rm{diag}}\{ I\}$ results in the expressions for the steady-state MSD as given by \eqref{fmsd}. Likewise, letting $\sigma_{k-1}=(I - \Pi _{k,1} )^{ - 1} \lambda_k$ results in the expressions for the steady-state EMSE as given by \eqref{femse}.

\end{document}